\documentclass[aps,prb,twocolumn,amsmath,amsfonts,showpacs,floatfix]{revtex4}
\usepackage{graphicx}

\begin{document}

\title{Interplay between direct and crossed Andreev reflections in hybrid nano-structures}

\author{Grzegorz Micha{\l}ek and Bogdan R. Bu{\l}ka}
\affiliation{Institute of Molecular Physics, Polish Academy of Science, ul. M. Smoluchowskiego 17, 60-179 Pozna\'n, Poland}
\author{Tadeusz Doma\'{n}ski and Karol I.\ Wysoki\'{n}ski}
\affiliation{Institute of Physics, M.\ Curie-Sk{\l}odowska University, pl. M. Curie-Sk{\l}odowskiej 1, 20-031 Lublin, Poland}

\date{\today}

\begin{abstract}
The interplay between various many body effects in a quantum dot attached to two normal and one superconducting lead is considered in the limit of large superconducting gap. By the proximity effect the superconducting lead induces pairing correlations on the quantum dot. In the subgap region one observes the anomalous tunneling via direct and crossed Andreev scattering, whereas the usual single particle electronic transfer is suppressed. The interactions of electrons on the dot leading to such phenomena as the Coulomb blockade and the Kondo effect severely modify the currents flowing in the system. In particular: (i) they prevent the existence of the negative differential conductance observed for non-interacting quantum dot over the whole range of voltages, (ii) affect the distribution of the currents as function of the applied voltage and (iii) lead to the appearance of additional low bias feature due to the formation of the Abrikosov-Suhl resonance. The non-local correlations in the Coulomb blockade regime are most pronounced for the particle-hole symmetric dot and thus can be easily tuned by means of gate voltage. They are observed even in the Kondo regime and dominate the behavior close to the Abrikosov-Suhl resonance showing convincingly that Kondo correlations do not destroy subtle entanglement between electrons.
\end{abstract}


\maketitle

\section{Introduction}

The hybrid multiterminal systems with a quantum dot and normal, superconducting and/or ferromagnetic electrodes are a source of rich physics~\cite{franceschi2010} with potentially interesting applications in spintronics~\cite{eschrig2011} or quantum information processing~\cite{burkard2000}. They allow the study of Andreev transport in the presence of Coulomb correlations~\cite{martinrodero2011}. One of the motivations is the possibility of producing entangled electrons resulting from splitting of Cooper pairs. This can be observed via non-local conductances due to the Andreev reflections.

In the paper [\onlinecite{recher2001}] it has been proposed to realize the goal using three terminal hybrid devices with quantum dots. On the other hand Ref. [\onlinecite{recher2002}] considered the quantum point contacts between the superconducting electrode and two Luttinger wires. The signatures of current correlations indicating the entanglement have been experimentally seen in devices with direct contact between two normal and one superconducting lead~\cite{beckmann2004, russo2005, wei2010} or with those where leads were contacted \textit{via} two or three quantum dots~\cite{hofstetter2009, hermann2010}. The multiterminal hybrid structures~\cite{hofstetter2009, hermann2010, lu2009, futterer2009, futterer2010, eldridge2010} are subject of recent studies, both theoretical~\cite{lambert1998, martinrodero2012, whan1996, levy1997, kang1998, fazio1998, schwab1999, raimondi1999, ivanov1999, sun1999, MKKIWActa, clerk2000, sun2000, avishai2001, krawiec2002, liu2004, bergeret2006, claughton1995, tanaka2007, andersen2011, hiltscher2011} and experimental~\cite{buitelaar2002, cleuziou2006, jarrillo-herrero2006, jorgensen2006, eichler2007, sand-jespersen2007, Deacon_etal, pillet2010, meshke2011}. The detailed understanding of these novel systems is very important as 'the effective use of the devices relies on the precise knowledge of the effects of interactions on the currents in the system'~\cite{schindele2012}.

In structures with quantum dot(s) and at least one superconducting lead one encounters various energy scales like temperature $T$, bias voltage $V$, superconducting gap $\Delta$, effective couplings $\Gamma$ between the quantum dot(s) and electrodes and charging energy $U$. Depending on their relation there exist various transport regimes. Of particular interest is the transport between the superconductor and the rest of the system. At bias voltages exceeding the superconducting gap or at high temperatures the single particle transport dominates, while for $V \ll \Delta$ the Andreev scattering~\cite{andreev1964} is the dominant transport mechanism.

The detailed analysis of the effect of Coulomb interactions on the dot on the Andreev transport in a three terminal device with single quantum dot is our primary goal here. We start with exactly solvable case of non-interacting dot and go through Coulomb blockade regime of transport ending up with Kondo correlated state. In the paper a three-terminal device (see Fig.~\ref{fig1}) with a superconducting electrode and two normal metallic electrodes connected via a quantum dot is considered. We assume that the superconducting gap is the largest energy scale. The Coulomb blockade is analyzed by means of Hubbard I approximation and we go beyond this approximation using equation of motion method~\cite{zubariev1960} and iterative perturbation theory~\cite{kajueter1996} (also known as modified second order perturbation theory~\cite{levy-yeyati1993}).

The superconducting correlations are induced in the quantum dot by the proximity effect to the superconducting lead. The Cooper pair injected from the superconducting lead to the dot either goes to one of the normal leads or splits and one of the electrons enters left (L) lead and other the right (R) one eventually retaining the singlet character of their state. In the reverse process an electron from a normal lead enters the superconductor leaving the hole behind in the same or other lead.

In three terminal device one distinguishes two different Andreev processes. In the direct Andreev reflection (DAR) two electrons entering the superconductor and the back-scattered hole are from the same lead while in crossed Andreev reflection (CAR) electrons stem from different normal leads. These non-local processes (CAR) are a potential source of entangled particles as they result from a singlet state of the Cooper pair. The processes competing with CAR are the single electron transfers (ET) between both normal electrodes. As the quantitative understanding of this competition is a prerequisite of the entangler based on quantum dot devices and the main goal of the paper, we shall quantify the competition by the non local differential conductance relating the current in the right lead flowing in response to the voltage in the left lead.

We have found that the Coulomb interactions generally suppress CAR processes in the large range of bias voltages. However, there remain regions in the vicinity of the Andreev bound states~\cite{meng2009} where the CAR processes dominate the transport and the total non-local conductance is negative indicating entanglement of pairs of separated electrons: one of them entering the left and the other one the right normal electrode. The most interesting finding is that these subtle quantum correlations are observed in the Kondo state, where the non-local conductance dominates over single electron transfer processes. This result agrees with the full counting statistics of two quantum dots in a three terminal device~\cite{soller2011}, which indicated the possibility of observing positive current cross-correlation in a Kondo regime of a hybrid structure. Our calculations have shown that the effect exists and we predict its observation in a device with a single quantum dot.

We note by passing that the related hybrid structures consisting of a quantum dot and one normal but two superconducting electrodes allow study of the interplay between the Josephson effect and Coulomb correlations~\cite{pala2007}. In a related work the systems similar to that studied here consisting of quantum dot contacted to normal, superconducting and ferromagnetic electrodes have been recently proposed to be an effective source of pure spin currents~\cite{lu2009, futterer2010}. The effect of non-collinear magnetization has also been discussed~\cite{zhu2001}.

The organization of the rest of the paper is as follows. In the next Section we present the model and approach to calculate currents flowing in the system under applied bias voltage. The differential conductances of the system with non-interacting quantum dot are calculated and discussed in Section III. The effect of electron interactions on the transport currents and conductances is studied in the Coulomb blockade regime (Section IV) and beyond it (Section V), using approximations which capture the Kondo correlations and are valid up to temperatures $T \approx T_K$. We end up with summary and conclusions.

\section{Description of the model and method of calculation}

\subsection{The hybrid device with a quantum dot}

\begin{figure}
\includegraphics[width=0.45\textwidth,clip]{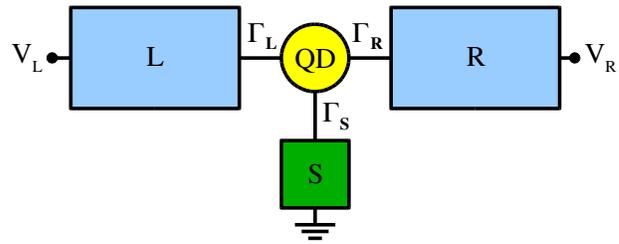}
\caption{(color online) Schematic view of a three-terminal device with a superconducting electrode (S) and two normal metallic electrodes (L, R) connected via a quantum dot (QD).}\label{fig1}
\end{figure}

We consider a system which consists of a quantum dot (QD) connected with two normal metal leads (the left - L and the right - R) and one superconducting (S) lead, see Fig.~\ref{fig1}. The system can be modeled by the Hamiltonian
\begin{equation}
\label{eq-Ham}
H = H_{QD} + \sum_{\alpha = L, R, S} H_\alpha + H_T \; ,
\end{equation}
where the first term describes the quantum dot, the second electrons in the leads and the third tunneling between the leads and the QD. The Hamiltonian of the QD reads
\begin{equation}
\label{eq-Ham-QD}
H_{QD} = \epsilon_0 \sum_\sigma d_\sigma^\dag d_\sigma + U n_\uparrow n_\downarrow \; ,
\end{equation}
where $\epsilon_0$ is the single-particle energy level, $d_\sigma^\dag$ ($d_\sigma$) denotes creation (annihilation) operator of the dot electron with spin $\sigma$, $n_\sigma \equiv d_\sigma^\dag d_\sigma$, and $U$ is the Coulomb interaction on QD. It is assumed that the normal metal electrodes are treated within the wide-band approximation
\begin{equation}
H_\alpha = \sum_{k, \sigma} \epsilon_{\alpha k} c_{\alpha k \sigma}^\dag c_{\alpha k \sigma} \; ,
\end{equation}
where $c_{\alpha k \sigma}^\dag$ ($c_{\alpha k \sigma}$) denotes creation (annihilation) of an electron with spin $\sigma$ and momentum $k$ in the electrode $\alpha = \{ L, R \}$. The third, superconducting electrode is described in the BCS approximation by
\begin{eqnarray}
H_S = \sum_{k, \sigma} \epsilon_{S k} c_{S k \sigma}^\dag c_{S k \sigma} \nonumber \\
+ \sum_k \left( \Delta c_{S -k \uparrow}^\dag c_{S k \downarrow}^\dag + \Delta^* c_{S k \downarrow} c_{S -k \uparrow} \right) \; ,
\end{eqnarray}
where we have assumed isotropic energy gap $\Delta$. Coupling between the QD and the external leads reads
\begin{equation}
H_T = \sum_{\alpha, k, \sigma} \left( t_\alpha c_{\alpha k \sigma}^\dag d_\sigma + t_\alpha^* d_\sigma^\dag c_{\alpha k \sigma} \right) \; ,
\end{equation}
where $t_\alpha$ is the hopping integral between QD and the $\alpha$ lead. An electron and hole transfer between the QD and the leads is described by an effective tunneling rate $\Gamma_\alpha$, which in the wide-band approximation takes the form $\Gamma_\alpha = 2 \pi \sum_k |t_\alpha|^2 \delta (E - \epsilon_{\alpha k}) = 2 \pi |t_\alpha|^2 \rho_\alpha$, where $\rho_\alpha$ is the density of states in the $\alpha$ electrode in the normal state.

The bias voltage $V_L$ ($V_R$) is applied to the left (right) electrode, while the superconducting electrode is grounded. Usually an additional gate is applied to the QD, by means of which one can change the position of the single-particle level $\epsilon_0$ and number of electrons $n$ on the dot.

\subsection{Currents and conductances}

The currents, which flow from the normal electrodes to the QD can be calculated from the time evolution of the total number operator~\cite{haug-jauho1996}
\begin{equation}
\label{curr}
I_\alpha \equiv - e \langle \dot{N}_\alpha \rangle = - \frac{i e}{\hbar} \langle [N_\alpha, H_T] \rangle \; .
\end{equation}
After standard manipulations (\ref{curr}) can be rewritten as
\begin{equation}
I_\alpha = \frac{4 e}{\hbar} \int \frac{d E}{2 \pi} \Gamma_\alpha \Im \left[ f_\alpha G_{11}^r + \frac{1}{2} G_{11}^< \right] \; ,
\end{equation}
where $G_{11}^r$ and $G_{11}^<$ are the matrix elements of the QD Green function $\hat{G}^r$ and $\hat{G}^<$ in the Nambu representation~\cite{sun2000}. Using the equation of motion technique (EOM) for the (non-equilibrium) Green function~\cite{haug-jauho1996, Domanski-08, niu1999, meir1991,sun1999} one can find currents originating from various types of tunneling processes. In actual calculations for an interacting system it is important to correctly determine the function $\hat{G}^<$ (and related local Wigner distribution function) for the nonequilibrium situation. For noninteracting case one can find exact expression for $\hat{G}^<$ (assuming quasielastic transport, for which the current conservation rule is fulfilled for any energy $E$). In the presence of interactions we use the relation $\hat{G}^< = \hat{G}^r \hat{\Sigma}^< \hat{G}^a$ and the ansatz proposed by Fazio and Raimondi~\cite{fazio1998} that the self-energies are proportional to that in the noninteracting case, $\hat{\Sigma}^{<, >} = \hat{\Sigma}_0^{<, >} \hat{A}$. The matrix $\hat{A}$ is determined by the condition $\hat{\Sigma}^< - \hat{\Sigma}^> = \hat{\Sigma}^r - \hat{\Sigma}^a$, which guarantees a current conservation.

In the subgap regime $|eV| < \Delta$ only the following components survive and the current can be expressed in terms of $G^r_{11}$ and $G^r_{12}$ components of the retarded Green function in Nambu space. Needless to say that in order to calculate $G^r_{11}$ and $G^r_{12}$ in the non-equilibrium system the full matrix Green function in Keldysh-Nambu space has to be calculated. The current flowing from the left-electrode reads
\begin{eqnarray}
\label{tot-cur-L}
I_L^{TOT} = I_L^{ET} + I_L^{AR} = I_L^{ET} + I_L^{DAR} + I_L^{CAR} \; ,
\end{eqnarray}
where (omitting energy E arguments)
\begin{eqnarray}
\label{ET-cur-L}
I_L^{ET} = \frac{2 e}{\hbar} \int \frac{dE}{2 \pi} \Gamma_L | G_{11}^r |^2 \Gamma_R ( f_L - f_R ) \; , \\
\label{DAR-cur-L}
I_L^{DAR} = \frac{2 e}{\hbar} \int \frac{dE}{2 \pi} \Gamma_L | G_{12}^r |^2 \Gamma_L ( f_L - \tilde{f}_L ) \; , \\
\label{CAR-cur-L}
I_L^{CAR} = \frac{2 e}{\hbar} \int \frac{dE}{2 \pi} \Gamma_L | G_{12}^r |^2 \Gamma_R ( f_L - \tilde{f}_R ) \; .
\end{eqnarray}
$f_\alpha \equiv f_\alpha (E) = \{ \exp[( E - e V_\alpha ) / k_B T] + 1 \}^{-1}$ and $\tilde{f}_\alpha \equiv \tilde{f}_\alpha (E) = 1 - f_\alpha(-E) = \{ \exp[( E + e V_\alpha ) / k_B T] + 1 \}^{-1}$ are the Fermi-Dirac distribution functions in the electrode $\alpha = \{ L, R \}$ for electrons and holes, respectively. Here, $I_L^{ET}$ denotes the current due to the normal electron transfer (ET) processes, while $I_L^{AR}$ is the Andreev current caused by the Andreev reflection (AR). The Andreev current can be divided into two parts: that due to the direct AR processes (DAR) and that due to the crossed AR processes (CAR). Similarly one can derive the current flowing from the R-electrode $I_R^{TOT}$ as well as from the S-electrode $I_S^{TOT}$ and check that the Kirchoff's law is fulfilled
\begin{eqnarray}
I_L^{TOT} + I_R^{TOT} + I_S^{TOT} = 0 \; .
\end{eqnarray}

For higher voltages, exceeding the energy gap $|eV| \geq \Delta$, there would be additional contributions to the electron transport, namely the single-particle tunneling $(2e / \hbar) \int (dE / 2 \pi)\; \Gamma_L | G_{11}^r |^2 \Gamma_S ( f_L - f_S )$ and the branch crossing processes $(2e / \hbar) \int (dE / 2 \pi)\; \Gamma_L | G_{12}^r |^2 \Gamma_S ( f_L - f_S )$. Let us note that with $\Gamma_S = 0$ the current in the superconducting electrode vanishes due to the fact that the Green function $G_{12}^r$ is proportional to $\Gamma_S$.

In a three terminal device one can define a non-local conductances i.e. related to the current flowing in the L (R) electrode due to the voltage applied to R (L) one. In accordance to the contributions $\kappa = \{ ET, DAR, CAR \}$ to the currents we shall also discuss the related conductances. Various differential conductances are defined as
\begin{equation}
\label{G_def}
\mathcal{G}_{\alpha / \beta}^\kappa = (-1)^{1 - \delta_{\alpha \beta}} \frac{d I_\alpha^\kappa}{d V_\beta} \; ,
\end{equation}
where $\alpha = \{ L, R, S \}$, $\beta = \{ L, R \}$, and $\delta_{\alpha \beta}$ is the Kronecker delta. Occasionally we shall also discuss the total conductances ($\kappa = TOT$) related to the total currents in a given lead.

\subsection{Green function of the quantum dot}

Eqs.~(\ref{tot-cur-L})-(\ref{CAR-cur-L}) show that to calculate currents flowing in the system one needs the full Green function $\hat{G}^r (E)$ of QD taking into account the Coulomb interactions and the couplings to the leads. From the Dyson equation
\begin{equation}
\label{Dyson}
\hat{G}^r (E) = \hat{g}^r (E) + \hat{g}^r (E) \hat{\Sigma}^r (E) \hat{G}^r (E) \; ,
\end{equation}
where $\hat{g}^r (E)$ is the Green function of the isolated or non-interacting dot and $\hat{\Sigma}^r (E)$ is the appropriate self-energy one can find that (omitting the energy argument $E$)
\begin{eqnarray}
\label{G11_U}
G_{11}^r = \frac{1 / g_{22}^r - \Sigma_{22}^r}{\left( 1 / g_{11}^r - \Sigma_{11}^r \right) \left( 1 / g_{22}^r - \Sigma_{22}^r \right) - \Sigma_{12}^r \Sigma_{21}^r} \; , \\
\label{G12_U}
G_{12}^r = - \frac{\Sigma_{12}^r}{1 / g_{22}^r - \Sigma_{22}^r} G_{11}^r \nonumber \\
= - \frac{\Sigma_{12}^r}{\left( 1 / g_{11}^r - \Sigma_{11}^r \right) \left( 1 / g_{22}^r - \Sigma_{22}^r \right) - \Sigma_{12}^r \Sigma_{21}^r} \; .
\end{eqnarray}

\section{Results for non-interacting quantum dot}

For the sake of later comparison we start the analysis with a simple example of non-interacting electrons $U = 0$ on the quantum dot, where analytical expressions can be found at $T = 0$. We discuss the density of states (DOS), and the conductances of the system. As our main focus is on the Andreev reflection processes we assume that transmission rates, the bias voltages and the temperature are much smaller than the energy gap of the superconducting electrode, i.e. $\Gamma_L$, $ \Gamma_R$, $\Gamma_S$, $eV_L$, $eV_R$, $k_B T \ll \Delta$. As already mentioned we assume the validity of these relations throughout the whole paper.

\subsection{Density of states}

For the study of non-interacting quantum dot we take the Green functions (\ref{G11_U}), (\ref{G12_U}) with a Green function for an isolated single-level QD
\begin{eqnarray}
\label{GF_T}
\hat{g}^r = \left(
\begin{array}{cc}
\displaystyle{\frac{1}{E - \epsilon_0 + i 0^+}} & 0 \\
0 & \displaystyle{\frac{1}{E + \epsilon_0 + i 0^+}}
\end{array} \right)
\end{eqnarray}
and self energies $\Sigma_{ij}^r$ ($i, j = \{ 1, 2 \}$) evaluated in the so called 'superconducting atomic limit' or deep inside the superconducting energy gap~\cite{tanaka2007}
\begin{equation}
\label{sigma0}
\hat{\Sigma}^r = \left(
\begin{array}{cc}
- i ( \Gamma_L + \Gamma_R ) / 2 & - \Gamma_S / 2 \\
- \Gamma_S / 2 & - i ( \Gamma_L + \Gamma_R ) / 2 \\
\end{array} \right) \; .
\end{equation}

It is an easy exercise to find the insightful expressions for matrix elements $G_{11}^r$ and $G_{12}^r$ of the retarded Green function valid in the limit of $\Delta \gg \Gamma_\alpha$ (i.e. for $\Delta \rightarrow \infty$)
\begin{eqnarray}
G_{11}^r = \frac{1}{2} \left( 1 + \frac{\epsilon_0}{E_d} \right) \frac{1}{E - E_d + i \Gamma_N / 2} \nonumber \\
+ \frac{1}{2} \left( 1 - \frac{\epsilon_0}{E_d} \right) \frac{1}{E + E_d + i \Gamma_N / 2} \;
\label{gf11}
\end{eqnarray}
and
\begin{eqnarray}
G_{12}^r = - \frac{\Gamma_S}{4 E_d} \frac{1}{E - E_d + i \Gamma_N / 2} \nonumber \\
+ \frac{\Gamma_S}{4 E_d} \frac{1}{E + E_d + i \Gamma_N / 2} \; .
\label{gf12}
\end{eqnarray}

For QD coupled to the superconducting lead, the proximity effect leads to the BCS-like structure of the spectral function and density of states $DOS = - \frac{1}{\pi} \Im{G_{11}^r}$ on QD given by
\begin{eqnarray}
DOS = \frac{1}{2 \pi} \left( 1 + \frac{\epsilon_0}{E_d} \right) \frac{\Gamma_N / 2}{( E - E_d )^2 + \Gamma_N^2 / 4} \nonumber \\
+ \frac{1}{2 \pi} \left( 1 - \frac{\epsilon_0}{E_d} \right) \frac{\Gamma_N / 2}{( E + E_d )^2 + \Gamma_N^2 / 4} \; .
\label{dos}
\end{eqnarray}
Density of states is a sum of the two Lorentzian curves centered at the $E = \pm E_d = \pm \sqrt{\epsilon_0^2 + \Gamma_S^2 / 4}$ and with the width of the peak $\Gamma_N / 2 = ( \Gamma_L + \Gamma_R ) / 2$. It means that in the QD, two Andreev bound states are formed: the "particle" state at $E = E_d$ and the "hole" state at $E = -E_d$, due to the proximity effect. For small $\Gamma_S \ll \Gamma_N$, the particle and hole peaks effectively merge into a single one at energy $E \approx \epsilon_0$. On the other hand for the strong coupling to the superconducting lead $\Gamma_S \gg \Gamma_N$ one observes in the DOS two separate peaks [with their weights depending on $\epsilon_0$ as visible from Eqs.~(\ref{gf11}) or (\ref{dos})] due to the proximity effect.

\subsection{Asymmetric bias}

With energy independent self-energies and for temperature $T = 0$ we find analytical formulas for the conductances. We show here the expression valid for the bias $eV_L$ applied to the left electrode, with R and S electrodes grounded ($eV_R = eV_S = 0$):
\begin{eqnarray}
\label{eq26}
\mathcal{G}_{L/L}^{ET}(eV_L) = \mathcal{G}_{R/L}^{ET}(eV_L) = \frac{4 e^2}{h} \frac{1}{2} \nonumber \\
\times \frac{\Gamma_L \Gamma_R \left[ ( \epsilon_0 + eV_L )^2 + \Gamma_N^2 / 4 \right]}{\left[ ( eV_L + E_d )^2 + \Gamma_N^2 / 4 \right] \left[ ( eV_L - E_d )^2 + \Gamma_N^2 / 4 \right]} \; , \\
\label{eq27}
\mathcal{G}_{L/L}^{DAR}(eV_L) = \frac{4 e^2}{h} \frac{1}{4} \nonumber \\
\times \frac{\Gamma_S^2 \Gamma_L^2}{\left[ ( eV_L + E_d )^2 + \Gamma_N^2 / 4 \right] \left[ ( eV_L - E_d )^2 + \Gamma_N^2 / 4 \right]} \; , \\
\label{eq28}
\mathcal{G}_{L/L}^{CAR}(eV_L) = - \mathcal{G}_{R/L}^{CAR}(eV_L) = \frac{4 e^2}{h} \frac{1}{8} \nonumber \\
\times \frac{\Gamma_S^2 \Gamma_L \Gamma_R}{\left[ ( eV_L + E_d )^2 + \Gamma_N^2 / 4 \right] \left[ ( eV_L - E_d )^2 + \Gamma_N^2 / 4 \right]} \; .
\end{eqnarray}
These formulas clearly demonstrate resonant transmission through two bound states $\pm E_d$.
\begin{figure}
\includegraphics[width=0.4\textwidth,clip]{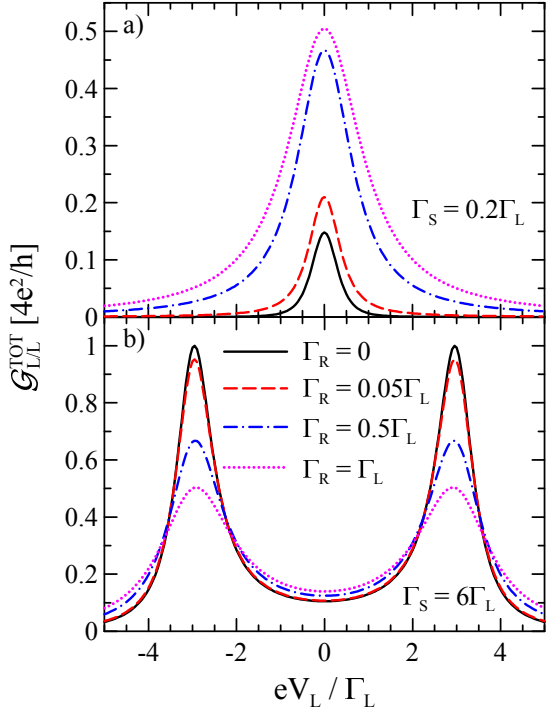}
\caption{(color online) Conductance $\mathcal{G}_{L/L}^{TOT}$ for (a) a week coupling $\Gamma_S = 0.2 \Gamma_L$ and (b) a strong coupling $\Gamma_S = 6 \Gamma_L$ to the S-electrode for various couplings $\Gamma_R = 0$ (black solid line), $\Gamma_R = 0.05 \Gamma_L$ (red dashed line), $\Gamma_R = 0.5 \Gamma_L$ (blue dash-dot line) and $\Gamma_R = \Gamma_L$ (magenta dotted line) at $\epsilon_0 = 0$ and for $V_R = V_S = 0$.}\label{fig2}
\end{figure}

In the three-terminal hybrid system various electronic transfer processes compete with each other. First of all, electron tunneling (ET) between normal electrodes competes with the Andreev reflection (AR). The non-local differential conductance $\mathcal{G}_{R/L}^{TOT} = \mathcal{G}_{R/L}^{ET} + \mathcal{G}_{R/L}^{CAR}$ (with the current measured at the right electrode as a response to voltage in the left one) can be positive when the ET processes dominate, or negative for a strong crossed Andreev reflection [compare Eqs.~(\ref{eq26}) and (\ref{eq28})]. For an asymmetric coupling $\Gamma_R > 2 \Gamma_L$ the CAR processes can dominate over the DAR processes [compare Eqs.~(\ref{eq27}) and (\ref{eq28})].

Fig.~\ref{fig2}a presents the total conductance $\mathcal{G}_{L/L}^{TOT}$ in the left junction in the case of weak coupling $\Gamma_S$ and for various couplings to the right electrode. In this case the particle-hole (p-h) splitting is not visible and $\mathcal{G}_{L/L}^{TOT}$ is dominated by ET processes. The conductance increases with an increase of $\Gamma_R$ and reaches maximum for symmetric coupling to the normal electrodes $\Gamma_L = \Gamma_R$. For larger $\Gamma_R > \Gamma_L$ the amplitude decreases. When $\epsilon_0 \neq 0$ the total conductance peaks are shifted and reduced.

For $\Gamma_S > \Gamma_N$ (Fig.~\ref{fig2}b) the p-h splitting is manifested in $\mathcal{G}_{L/L}^{TOT}$ as two peaks centered at $eV_L = \pm E_d$. Now, the proximity effect is strong and the AR processes are relevant. The amplitude of the conductance always decreases with $\Gamma_R$. From the formulas (\ref{eq27}) and (\ref{eq28}) one can find that when the $\Gamma_R < 2 \Gamma_L$ the contribution from the CAR processes is smaller than that one from the DAR processes. On the other hand, the CAR processes contribute to the conductance more effectively than the DAR processes when $\Gamma_R > 2 \Gamma_L$. The relative hight of the total conductance peaks changes also in a different way with $\epsilon_0$. That around $eV_L = -E_d$ for large values of $\epsilon_0$ changes like $(2e^2 / h) \gamma^2 / (\epsilon_0^2 + \gamma^2)$, with some effective
coupling $\gamma$, while the peak around $eV_L = E_d$ saturates in this limit at the value $(2e^2 / h)$.

\subsection{Competition between ET and CAR - negative conductance}

\begin{figure}
\includegraphics[width=0.4\textwidth,clip]{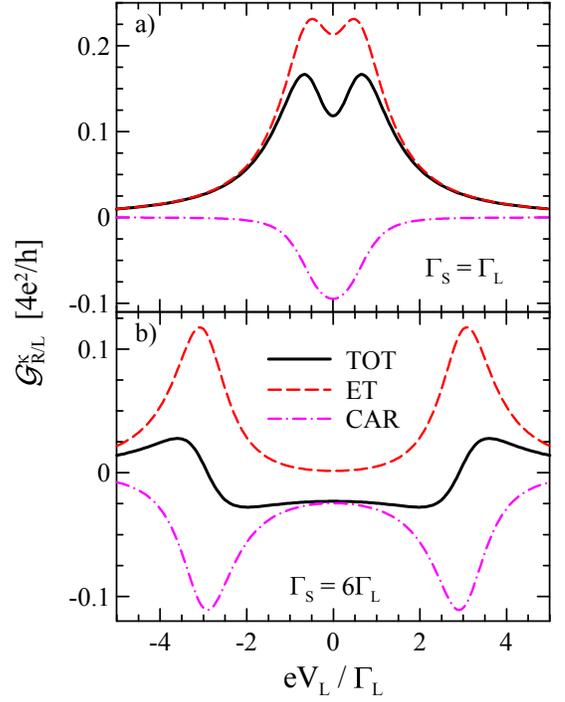}
\caption{(color online) Conductance $\mathcal{G}_{R/L}^{TOT}$ (black solid line) measured at the R-electrode with respect to the potential $eV_L$ applied to the L-electrode for (a) $\Gamma_S = \Gamma_L$ and (b) $\Gamma_S = 6 \Gamma_L$. $\mathcal{G}_{R/L}^{ET}$ (red dashed line) and $\mathcal{G}_{R/L}^{CAR}$ (magenta dash-dot line) present the conductance contributions due to the direct electron transfer and CAR processes, respectively. The other parameters are $\Gamma_R = 0.5 \Gamma_L$, $\epsilon_0 = 0$ and $V_R = V_S = 0$.}\label{fig3}
\end{figure}

The non-local conductance $\mathcal{G}^{CAR}_{R/L}$ is a direct measure of the entangled current. As mentioned the competing process is that due to direct electron transfer between normal electrodes. The results for $\mathcal{G}_{R/L}^\kappa$ are presented in Fig.~\ref{fig3}. The total conductance $\mathcal{G}_{R/L}^{TOT}$ in the R-junction has only two components: normal ET, which is always positive, while the CAR processes give negative contribution to the total conductance. As long as the $\Gamma_S < \Gamma_N$ the ET contribution is larger than the CAR contribution and the $\mathcal{G}_{R/L}^{TOT}$ is positive, see Fig.~\ref{fig3}a. However, in the opposite case $\Gamma_S > \Gamma_N$, the $\mathcal{G}_{R/L}^{TOT}$ can be negative, because the CAR processes dominate over the direct electron tunneling (ET). For the symmetric case (with $\epsilon_0 = 0$) $\mathcal{G}_{R/L}^{TOT}$ is negative between the Andreev bound states. When the gate voltage is applied to the QD ($\epsilon_0 \neq 0$) the electron-hole symmetry is broken and the $\mathcal{G}_{R/L}^{TOT}$ characteristics are asymmetric with respect to $eV_L = 0$. This behavior is caused by the ET contribution, which amplitude depends on the position of $\epsilon_0$ [see numerator of Eq.~(\ref{eq26})]. Now, the ET contribution prefers the hole (electron) resonance level $-E_d$ ($+E_d$) for $\epsilon_0 < 0$ ($\epsilon_0 > 0)$. On the other hand $\mathcal{G}_{R/L}^{CAR}$ is always symmetric with respect to $\epsilon_0 = 0$ and $eV_L =0$, see Eq.~(\ref{eq28}).

The dominance of the CAR over ET processes in the non-local conductance $\mathcal{G}_{R/L}$ requires $\Gamma_S > \Gamma_N$ and is visible for the voltages $eV_L$ fulfilling
\begin{equation}
|\epsilon_0 + eV_L| \le \sqrt{\Gamma_S^2 - \Gamma_N^2} \; ,
\label{car-et}
\end{equation}
as it can be easily deduced from equations (\ref{eq26}) and (\ref{eq28}). In other words, CAR processes dominate for the voltages $eV_L$ for which the anomalous self-energy ($\Gamma_S / 2$ in the non-interacting case) dominates nominator of the $G^r_{11}$ Green function.

\section{Effect of Coulomb blockade}

The non-interacting quantum dot in contact with superconductor develops two Andreev bound states at $\pm \sqrt{\epsilon_0^2 + \Gamma_S^2 / 4}$ and the non-local conductance is dominated by the Copper pair splitting processes for the voltage $-E_d<eV_L<E_d$. With the Coulomb interaction taken into account the exact solution is no more available and the approximations are necessary. In order to gain some insight into the effect of correlations we shall use the formally exact expression for the Green functions (\ref{Dyson}) and calculate the self-energies approximately. We again assume that the superconducting order parameter $\Delta$ is the largest energy scale and calculate the contributions to the leads induced self-energy to lowest order in the coupling getting Eq.~(\ref{sigma0}). The contribution of Coulomb interactions to the self-energy will be calculated in Hubbard I approximation~\cite{HubbardI}, equation of motion (EOM) and iterative perturbation approach (IPT). Since we consider the paramagnetic case $\langle n_\uparrow \rangle = \langle n_\downarrow \rangle = n / 2$, the total accumulated charge $n$ at QD (required to get correct value of Coulomb self energy) is calculated in the self-consistent way from the equation
\begin{equation}\label{nU1}
n = 2 \int \frac{dE}{2 \pi i} G_{11}^< (E) \; .
\end{equation}
The lesser Green function
\begin{eqnarray}
\label{lesserG}
G_{11}^< = i | G_{11}^r |^2 ( \Gamma_L f_L + \Gamma_R f_R ) \nonumber \\
+ i | G_{12}^r |^2 ( \Gamma_L \tilde{f}_L + \Gamma_R \tilde{f}_R ) \;
\end{eqnarray}
is calculated using the Green functions (\ref{G11_U}), (\ref{G12_U}) with the following Green function for an isolated single-level QD in presence of the Coulomb interactions~\cite{haug-jauho1996}
\begin{eqnarray}
\label{HubbardI}
g_{11}^r = \frac{1 - \langle n_\downarrow \rangle}{E - \epsilon_0 + i 0^+} + \frac{\langle n_\downarrow \rangle}{E - \epsilon_0 - U + i 0^+} \; , \nonumber \\
g_{22}^r = \frac{1 - \langle n_\uparrow \rangle}{E + \epsilon_0 + i 0^+} + \frac{\langle n_\uparrow \rangle}{E + \epsilon_0 + U + i 0^+} \; .
\end{eqnarray}
The local current conservation rule is fulfilled within this approximation and one can describe the Coulomb blockade effect in transport through QDs. The approximation neglects, however, spin-flip processes in tunneling and ignores the Kondo correlations so it can be applied for high temperatures (well above the Kondo temperature $T_K$). Equation of motion and IPT techniques allow to go beyond Coulomb blockade and will be considered in the next section.

\subsection{Density of states modified by Coulomb interactions}

\begin{figure}
\includegraphics[width=0.4\textwidth, clip]{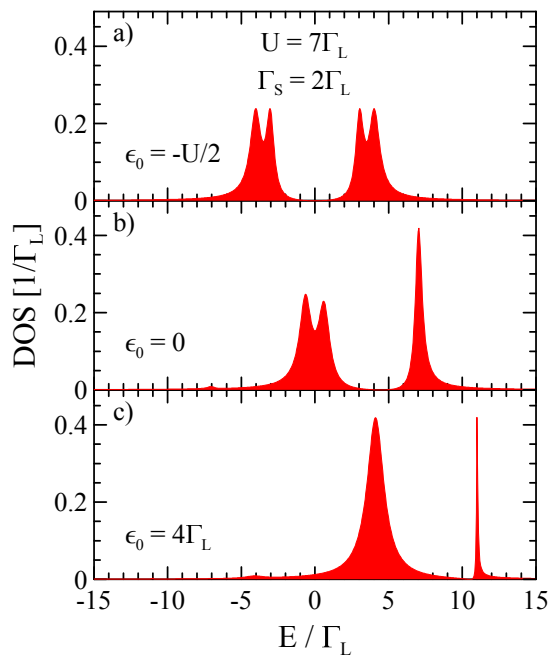}
\caption{(color online) The equilibrium DOS for a large Coulomb interaction $U = 7 \Gamma_L$ and $\Gamma_S = 2 \Gamma_L$, and for (a) $\epsilon_0 = -U/2$ (the electron-hole symmetry point), (b) $\epsilon_0 = 0$ (the end of the Coulomb blockade range) and (c) $\epsilon_0 = 4 \Gamma_L$ (the empty dot regime) with $\Gamma_R = 0.5 \Gamma_L$.}\label{fig4}
\end{figure}

We start by presenting numerical results for the density of states at equilibrium. To simplify calculations we assume temperature $T = 0$. For finite $T > 0$ one has to include also thermal broadening in all plots presented below, but the physics is the same.

The density of states of the interacting quantum dot shows four peaks. With Coulomb interaction $U$ the Green function $G^r_{11}(E)$ [Eq.~(\ref{G11_U})] has four poles and the spectrum consists of four Andreev bound states. This is related to splitting of the dot spectrum into lower and upper Hubbard level and the mixing of empty and doubly occupied states. For the particle-hole symmetric case, $\epsilon_0 = -U/2$, all peaks have the same amplitude (see Fig.~\ref{fig4}a). For a large Coulomb interaction two pairs of Andreev peaks are separated by a wide Coulomb blockade region. The gate voltage can be used to tune the positions and the amplitude of the Andreev peaks. At $\epsilon_0 = 0$ one reaches the end of the Coulomb blockade region. DOS becomes asymmetric and dependent on the electron concentration $n$ (see Fig.~\ref{fig4}b). With a further increase of $\epsilon_0$ the system goes to the empty dot regime, in which only two right most peaks survive (see the plot in the Fig.~\ref{fig4}c). The inner peak has a Lorentzian shape, while the outer one is very narrow and asymmetric. Moreover DOS reaches zero between the peaks. This indicates the Fano resonance and destructive interference of waves scattered on the Andreev bound states.

The positions of the Andreev bound states can be found from poles of the Green function $G_{11}^r$. In the limit $\Gamma_L$, $\Gamma_R \rightarrow 0$ one gets an analytical expression
\begin{equation}
E_{\lambda, \lambda'}^A = \frac{\lambda}{\sqrt{2}} \sqrt{\epsilon_0^2 + \epsilon_U^2 + \Gamma_S^2 / 4 + \lambda' \delta}
\label{bound_states}
\end{equation}
where $\delta = \sqrt{( \epsilon_0^2 + \epsilon_U^2 + \Gamma_S^2 / 4 )^2 - ( \Gamma_S^2 \epsilon_n^2 + 4 \epsilon_0^2 \epsilon_U^2 )}$, $\epsilon_U = \epsilon_0 + U$, $\epsilon_n = \epsilon_0 + ( 1 - n/2 ) U$ and $\lambda, \lambda' = \pm 1$.

In the double occupancy regime (for $n \rightarrow 2$) one finds the inner peaks at $E_{\pm, -}^A = \pm \sqrt{( \epsilon_0 + U )^2 + \Gamma_S^2 / 4}$ and the outer peaks at $E_{\pm, +}^A = \pm |\epsilon_0|$. Similarly for $n \rightarrow 0$ (the empty dot regime) $E_{\pm, -}^A = \pm \sqrt{\epsilon_0^2 + \Gamma_S^2 / 4}$ and $E_{\pm, +}^A = \pm |\epsilon_0 + U|$. The height of the DOS peaks changes non-monotonically. For example in the empty dot regime the states $E_{+, -}^A$ and $E_{+, +}^A$ survive and they have the same height while the peaks corresponding to the states $E_{-, -}^A$ and $E_{-, +}^A$ are suppressed to zero. Moreover, in the empty dot regime, the width of the peak at $E_{+, +}^A$ goes to zero, while the peak at $E_{+, -}^A$ has the width $\Gamma_N / 2$ - the same value as for the noninteracting electrons. In the Coulomb blockade region $\epsilon_0 \in [-U, 0]$ the spectrum $E_{\lambda, \lambda'}^A$ is hybridized. The DOS peaks show strong changes going between different branches of $E_{\lambda, \lambda'}^A$.

\subsection{Linear transport}

\begin{figure}
\includegraphics[width=0.4\textwidth, clip]{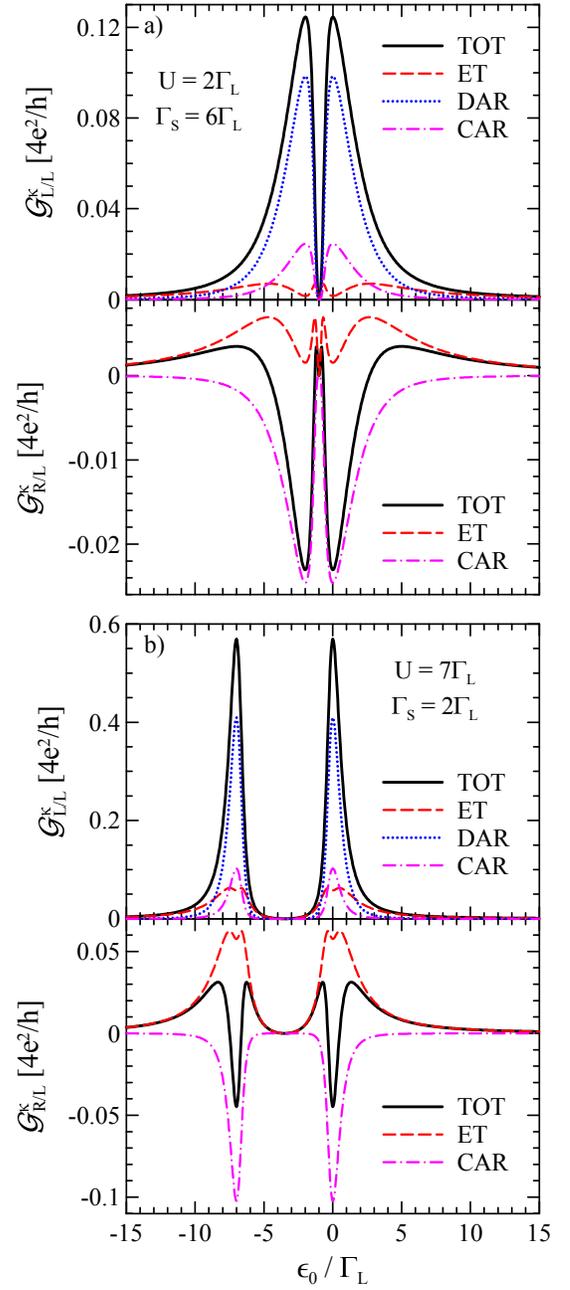}
\caption{(color online) Characteristics of conductance in the linear response regime, i.e. $V_L, V_R, V_S \rightarrow 0$. Top panel: $\mathcal{G}_{L/L}^{TOT} (0)$ (black solid line) with its components: $\mathcal{G}_{L/L}^{ET} (0)$ (red dashed line), $\mathcal{G}_{L/L}^{DAR} (0)$ (blue dotted line) and $\mathcal{G}_{L/L}^{CAR} (0)$ (magenta dash-dot line); bottom panel: $\mathcal{G}_{R/L}^{TOT} (0)$ (black solid line) with its components $\mathcal{G}_{R/L}^{ET} (0)$ (red dashed line) and $\mathcal{G}_{R/L}^{CAR} (0)$ (magenta dash-dot line). The results for (a) a small Coulomb interaction $U = 2 \Gamma_L$ and $\Gamma_S = 6 \Gamma_L$; and (b) a large Coulomb interaction $U = 7 \Gamma_L$ and $\Gamma_S = 2 \Gamma_L$ for the asymmetric coupling to the left and right electrode, $\Gamma_R = 0.5 \Gamma_L$.}\label{fig5}
\end{figure}

Here we study the influence of Coulomb interactions on the transport characteristics obtained in the linear regime, i.e. in the limit of a small bias voltage $V_\alpha \to 0$.

Results of the gate voltage dependence of the local and non-local conductances are presented in Fig.~\ref{fig5} for a small ($U < \Gamma_S$) and large ($U > \Gamma_S$) Coulomb interaction. The total conductances $\mathcal{G}_{L/L}^{TOT} (0)$ and $\mathcal{G}_{R/L}^{TOT} (0)$ as well as their components show particle-hole symmetry. The relative importance of the CAR and ET contributions to the linear conductances can be tuned by the gate voltage. The conductance $\mathcal{G}_{L/L}^{TOT} (0)$ has two well separated peaks at the ends of the Coulomb blockade region, i.e. close to $\epsilon_0 \approx -U$ and $\epsilon_0 \approx 0$. The main contribution to the conductance presented in Fig.~\ref{fig5} comes from the Andreev reflection processes, because the proximity effect is large ($\Gamma_S > \Gamma_N$). The behavior of $\mathcal{G}_{R/L}^{TOT} (0)$ is shown in the bottom panels in Fig.~\ref{fig5}. Again, in the close analogy to the non-interacting case one can see competition between the Andreev reflection and the direct electron transfer processes. As a result the conductance $\mathcal{G}_{R/L}^{TOT} (0)$ can be negative. However, in contrast to the non-interacting case, when the conductance $\mathcal{G}_{R/L}^{TOT} (0) < 0$ in the whole region between the Andreev bound state (see Fig.~\ref{fig3}b), now we observe $\mathcal{G}_{R/L}^{TOT} (0) > 0$ inside this region and it becomes negative ($\mathcal{G}_{R/L}^{TOT} (0) < 0$) in the vicinity of resonant levels. This is manifestation of the Coulomb blockade effect, which suppresses stronger the Andreev reflection processes than the direct electron transfers (compare the components $\mathcal{G}_{R/L}^{ET} (0)$ and $\mathcal{G}_{R/L}^{CAR} (0)$ in Fig.~\ref{fig5}). It is worth noting that the signatures of the four Andreev bound states are only visible in the ET components of both local and non-local conductances in the linear regime.

\subsection{Nonlinear transport characteristics}

\begin{figure}
\includegraphics[width=0.4\textwidth,clip]{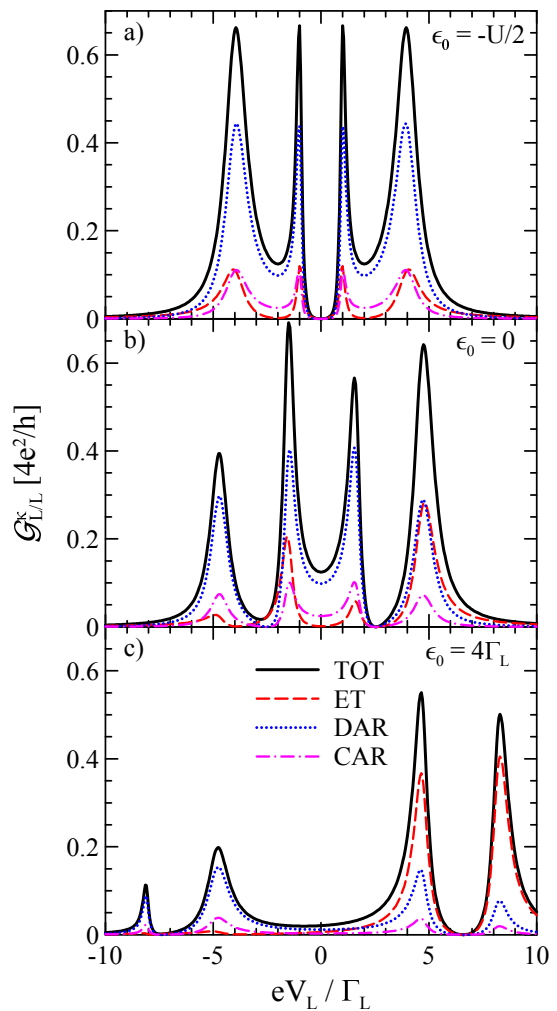}
\caption{(color online) Voltage dependence of conductance $\mathcal{G}_{L/L}^{TOT}$ (black solid line) with its components: $\mathcal{G}_{L/L}^{ET}$ (red dashed line), $\mathcal{G}_{L/L}^{DAR}$ (blue dotted line) and $\mathcal{G}_{L/L}^{CAR}$ (magenta dash-dot line) for (a) $\epsilon_0 = -U/2$, (b) $\epsilon_0 = 0$ and (c) $\epsilon_0 = 4 \Gamma_L$. The other parameters are $V_R = V_S = 0$, $U = 4 \Gamma_L$, $\Gamma_S = 6 \Gamma_L$ and $\Gamma_R = 0.5 \Gamma_L$.}\label{fig6}
\end{figure}

\begin{figure}
\includegraphics[width=0.4\textwidth,clip]{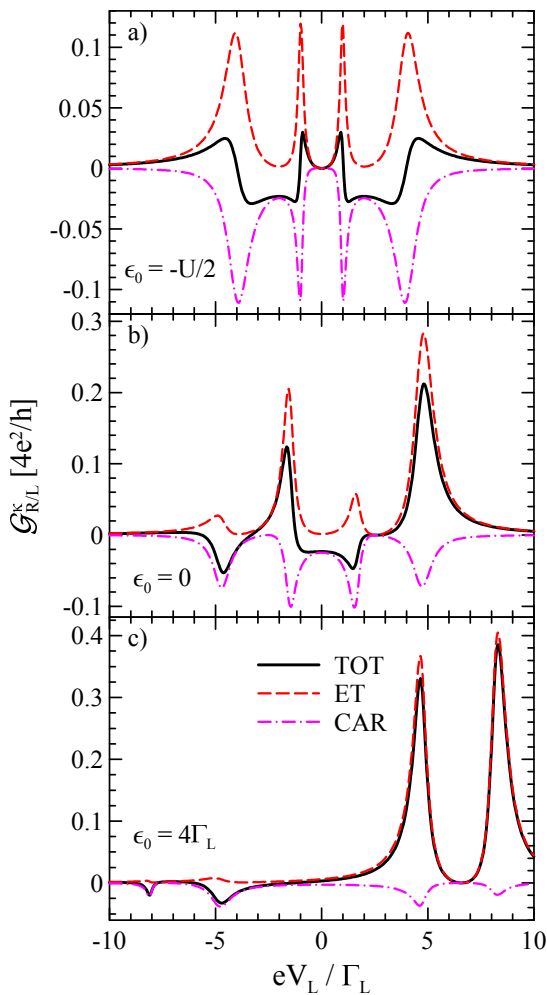}
\caption{(color online) Voltage dependence of the conductance $\mathcal{G}_{R/L}^{TOT}$ (black solid line) with its components: $\mathcal{G}_{R/L}^{ET}$ (red dashed line) and $\mathcal{G}_{R/L}^{CAR}$ (magenta dash-dot line) for (a) $\epsilon_0 = -U/2$, (b) $\epsilon_0 = 0$ and (c) $\epsilon_0 = 4 \Gamma_L$. The other parameters are $V_R = V_S = 0$, $U = 4 \Gamma_L$, $\Gamma_S = 6 \Gamma_L$ and $\Gamma_R = 0.5 \Gamma_L$.}\label{fig7}
\end{figure}

Outside the linear voltage regime we calculate currents and differential conductances taking full voltage dependence of the Fermi functions in the current formulas.

Fig.~\ref{fig6} and Fig.~\ref{fig7} present the conductance $\mathcal{G}_{L/L}^{TOT}$ and $\mathcal{G}_{R/L}^{TOT}$ as a function of the bias $V_L$. Comparing with the noninteracting case (Fig.~\ref{fig2}), in the presence of Coulomb interactions two additional conductance peaks appeared (Fig.~\ref{fig6}a), which correspond to Coulomb excitations. One sees the Coulomb blockade valley between them: the conductance $\mathcal{G}_{L/L}^{TOT}$ and all its components $\mathcal{G}_{R/L}^{ET}$, $\mathcal{G}_{L/L}^{DAR}$, $\mathcal{G}_{L/L}^{CAR}$ are reduced to zero in this region. The main contribution to the conductance $\mathcal{G}_{L/L}^{TOT}$ is from the DAR processes (see the blue dotted curve corresponding to $\mathcal{G}_{L/L}^{DAR}$ in Fig.~\ref{fig6}a). Figs.~\ref{fig6}a,~\ref{fig6}b and \ref{fig6}c present evolution of the conductance characteristics when the system goes to the empty dot regime. Notice that the asymmetry in the total conductance characteristics $\mathcal{G}_{L/L}^{TOT}$ is due to the ET contribution $\mathcal{G}_{L/L}^{ET}$, because $\mathcal{G}_{L/L}^{DAR}$ and $\mathcal{G}_{L/L}^{CAR}$ are almost symmetric with respect to $V_L = 0$. Moreover in the empty dot regime the ET contribution is enhanced, the AR processes are weakened. For $\epsilon_0 = 4 \Gamma_L$ the DAR processes dominate for $V_L < 0$ (see the blue dotted curve), whereas for $V_L > 0$ the ET tunneling plays an important role (the red dashed curve). Conductance and its components are strongly suppressed between two right peaks, what suggest dynamical Coulomb blockade. In this range the current is dynamically blocked for short time intervals, when an electron occupies the quantum dot.

The competition between the ET and CAR processes is well seen in the Fig.~\ref{fig7} presenting the conductance $\mathcal{G}_{R/L}^{TOT}$ determined on the R-junction. For the symmetric case $\epsilon_0 = -U/2$ the CAR processes are more strongly suppressed than the ET tunneling in the Coulomb blockade regime, and therefore, the total conductance $\mathcal{G}_{R/L}^{TOT}$ becomes positive. A similar effect one observes in Fig.~\ref{fig7}c in the dynamical Coulomb blockade region between two right peaks. Fig.~\ref{fig7}b presents the intermediate case $\epsilon_0 = 0$, where one can see how the Andreev bound states changed their role and how the ET and CAR processes compete with each other.

In the interacting case the non-local conductances are given by the energy integrals of the modules squared of $G_{11}^r (E)$ and $G_{12}^r (E)$ elements of the Green function for ET and CAR components, respectively (cf. integrals in Eqs. (\ref{ET-cur-L}) and (\ref{CAR-cur-L})). From the first equality in the formula (\ref{G12_U}) relating both components of the matrix Green function it follows that the contribution to CAR processes will dominate if
\begin{equation}
\left|\frac{\Sigma^r_{12}}{1/g^r_{22}-\Sigma^r_{22}}\right|^2>1
\label{factor}
\end{equation}
over the energy region; $0 < E < eV_L$ at $T = 0 K$. This mainly happens close to the Andreev resonances, when the denominator in (\ref{factor}) is small in comparison to the anomalous self-energy. This condition is general; the energy dependencies of the normal and anomalous self-energies over the integration range decide whether the CAR or ET processes dominate. The CAR component of the conductance show electron-hole symmetry with four Lorentzian resonance peaks around the Fermi energy $E_F = 0$. In contrast, the ET component have non-Lorentzian peaks, because an electron channel is preferred for transmission that leads to asymmetry well seen in Fig.~\ref{fig6}b, c and Fig.~\ref{fig7}b, c.

\section{Beyond Coulomb blockade: Kondo correlations}

From the physical point of view the Coulomb repulsion $U$ is responsible for the charging effect and, at lower temperatures, for the Kondo effect i.e. formation of the singlet resonant state between the spin localized on a QD and spins of itinerant electrons~\cite{hewson1997} from the normal leads. These effects spectroscopically manifest themselves by the appearance of the peaks around $E = \epsilon_0$ and $E = \epsilon_0 + U$ and the Kondo (or Abrikosov-Suhl) resonance in the density of states at the Fermi energy of the normal lead~\cite{Glazman, NgLee}. The width of the resonance is a characteristic scale, which is the Kondo temperature $T_K$. To estimate its value for a given set of parameters we use the formula~\cite{hewson1997}
\begin{equation}
k_BT_K = \sqrt{U \Gamma_N} \exp{\left[ \frac{\pi}{2} \frac{\epsilon_0 ( \epsilon_0 + U )}{U \Gamma_N} \right]} \; .
\label{tkondo}
\end{equation}
In non-equilibrium transport \textit{via} a quantum dot attached to two external electrodes two such resonances appear at the positions corresponding to the chemical potentials in the biased system~\cite{meir1991}. If the quantum dot is also coupled to the superconducting electrode the competition is observed~\cite{fazio1998} between the above mentioned features and the proximity induced on-dot-pairing.

To analyze the competition between currents beyond the Coulomb blockade limit we treat the electron interactions using the equation of motion (EOM) procedure~\cite{haug-jauho1996} and iterative perturbation theory (IPT)~\cite{martinrodero2011}. Both techniques have been previously used for studying interacting quantum dots in different setups~\cite{Domanski-08, Yamada-11, levy-yeyati1993}.

The equation of motion approach~\cite{zubariev1960}, which in general~\cite{kashcheyevs2006} \textit{"can form a basis for a qualitative analytic treatment of the Kondo effect"} is probably one of the simplest methods, qualitatively capturing~\cite{galperin2007} the physics of the non-equilibrium Kondo correlations at arbitrary $U$. The results, however, are not reliable on a quantitative level because of poor resolution of the Kondo peak. The comparison of the results obtained by EOM and the non-crossing approximation (NCA) shows~\cite{krawiec2002a} that the positions of the Kondo resonances are well described for system out of equilibrium. However, the method badly reproduces the half-filled situation (even on a qualitative level). For this reason we shall complementary use the iterative perturbation approach which is known to give correct results at half filling~\cite{kajueter1996} and has been adopted to the non-equilibrium transport via quantum dots~\cite{martinrodero2011, Yamada-11, levy-yeyati1993}.

To capture the Kondo physics we use the Dyson equation (\ref{Dyson}) with the noninteracting Green function (\ref{GF_T}) and impose the matrix self-energy $\hat{\Sigma}^{r, U} (E)$ in the following diagonal form
\begin{eqnarray}
\hat{\Sigma}^{r, U} (E) \simeq \left(
\begin{array}{cc}
\Sigma_N (E) & 0 \\
0 & - \left[ \Sigma_N (-E) \right]^*
\end{array} \right) \; .
\label{U_correction}
\end{eqnarray}
Within EOM approach the self-energy $\Sigma_N (E)$ reads~\cite{Domanski-08} (omitting the energy argument $E$)
\begin{eqnarray}
\Sigma_N = E - \epsilon_0 - \; \\
\frac{[E - \epsilon_0 - \Sigma_0][E - \epsilon_0 - \Sigma_0 - U - \Sigma_3] + U \Sigma_1}
{E - \epsilon_0 - \Sigma_0 - [\Sigma_3 + U (1 - \langle n_\downarrow \rangle)]} \; , \nonumber
\label{ansatz}
\end{eqnarray}
where~\cite{haug-jauho1996}

\begin{eqnarray}
\Sigma_0 = \sum_{\alpha = L, R} \sum_k \frac{|t_\alpha|^2}{E - \xi_{ \alpha k}} \simeq \frac{-i}{2} \left( \Gamma_L + \Gamma_R \right) \; , \\
\Sigma_\nu = \sum_{\alpha = L, R} \sum_k \left[ \frac{|t_{\alpha}|^2}{E - \xi_{ \alpha k}} + \frac{|t_{\alpha}|^2}{E - U - 2 \epsilon_0 + \xi_{\alpha k}} \right] \nonumber \\
\times \left\{
\begin{array}{ccc}
f(\xi_{\alpha k}) & \mathrm{for} & \nu = 1 \\
1 & \mathrm{for} & \nu = 3
\end{array} \right.
\label{sigmas}
\end{eqnarray}
and $\xi_{\alpha k} = \epsilon_{\alpha k} - eV_\alpha$.

The diagonal form of the self-energy (\ref{U_correction}) neglects any influence of the correlations $U$ on the induced on-dot pairing. Such an approximation has been shown~\cite{Domanski-08} to give a qualitative agreement with the experimental data obtained for InAs quantum dots~\cite{Deacon_etal}. Approximation (\ref{U_correction}) provides some insight into the physics of the hybrid structures discussed in this work but other advanced techniques~\cite{sellier2005} would be needed to describe an interplay between the Kondo and Andreev effects~\cite{martinrodero2011} on a some qualitative level. Analysis of the Kondo correlations~\cite{glazman1989, Domanski-08} under the non-equilibrium conditions~\cite{munoz2013} can be done, for instance using the suitably generalized non-crossing approximation~\cite{sellier2005, krawiec2002a} or the numerical renormalization group approach~\cite{tanaka2007, Bauer_2013}.

Let us recall~\cite{Domanski-08} that within EOM method the optimal conditions for enhancing the Andreev conductance by the Kondo resonance occur when $\Gamma_S \sim \Gamma_L$. One notices that the couplings to the normal electrodes $\Gamma_R$ and $\Gamma_L$ control the broadening of the quasi particle peaks at $\epsilon_0$ and $\epsilon_0 + U$. It means that for $\Gamma_S \sim \Gamma_L$ the particle-hole splitting is not well pronounced in the single particle spectrum in comparison to the results discussed in sections III and IV.

\begin{figure}
\includegraphics[width=0.4\textwidth, clip]{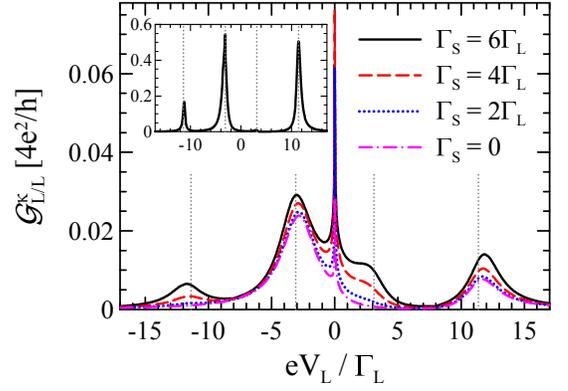}
\caption{(color online) Voltage dependence of conductance $\mathcal{G}_{L/L}^{TOT}$ obtained at low temperature $k_B T = 0.01 \Gamma_L$ in the Kondo regime for $\Gamma_S = 6 \Gamma_L$ (black solid line), $\Gamma_S = 4 \Gamma_L$ (red dashed line), $\Gamma_S = 2 \Gamma_L$ (blue dotted line) and $\Gamma_S = 0$ (magenta dash-dot line). The other parameters are $U = 14 \Gamma_L$, $\epsilon_0 = -3 \Gamma_L$, $\Gamma_R = 0.5 \Gamma_L$ and $V_R = V_S = 0$. The estimated $k_B T_K \approx 0.39 \Gamma_L$. The dotted vertical lines show the positions of subgap Andreev bound states for the case $\Gamma_S = 6 \Gamma_L$. Inset shows the results obtained within Hubbard I approximation for $\Gamma_S = 6 \Gamma_L$.}
\label{fig8}
\end{figure}

Fig.~\ref{fig8} shows the total differential conductance measured in the left lead for various couplings to the superconducting electrode. In the calculations we have assumed low temperatures and large $U = 14 \Gamma_L$ value to get all peaks separated and well developed Kondo resonance. For $\Gamma_S = 0$ we have two broadened resonant levels at $\epsilon_0$ and $\epsilon_0 + U$ and the Kondo peak appearing at the $eV_L = 0$. The zero bias resonance is due to the Abrikosov-Suhl resonances which appear at the Fermi levels of normal leads. Increasing coupling to the superconducting lead results in the four broadened Andreev states. The dotted vertical lines in the figure show the positions of the Andreev bound states calculated from Eqs. (\ref{bound_states}) for $\Gamma_S = 6 \Gamma_L$. The central peak corresponding to the Kondo resonance is observed for all values of coupling to the superconducting electrode. As already mentioned this feature has been recently observed experimentally~\cite{Deacon_etal} in the two terminal quantum dot. In the inset we show total (local) conductance obtained within Hubbard I approximation for $\Gamma_S = 6 \Gamma_L$. Note the nearly complete disappearance of one of the Andreev peaks in the Coulomb blockade regime and its partial recovery as well as the appearance of zero bias anomaly when Kondo correlations are taken into account (the main Figure).

\begin{figure}
\includegraphics[width=0.4\textwidth, clip]{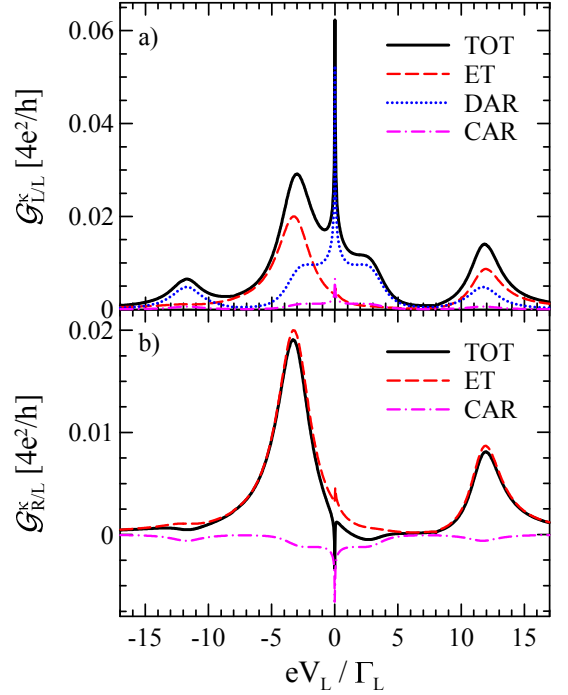}
\caption{(color online) Characteristics of conductance in the Kondo regime. (a) $\mathcal{G}_{L/L}^{TOT}$ (black solid line) with its components: $\mathcal{G}_{L/L}^{ET}$ (red dashed line), $\mathcal{G}_{L/L}^{DAR}$ (blue dotted line) and $\mathcal{G}_{L/L}^{CAR}$ (magenta dash-dot line); (b) $\mathcal{G}_{R/L}^{TOT}$ (black solid line) with its components $\mathcal{G}_{R/L}^{ET}$ (red dashed line) and $\mathcal{G}_{R/L}^{CAR}$ (magenta dash-dot line). The results are obtained for $k_B T = 0.01 \Gamma_L$ (i.e. well below $k_B T_K \approx 0.39 \Gamma_L$) using the model parameters $U = 14 \Gamma_L$, $\epsilon_0 = -3 \Gamma_L$, $\Gamma_R = 0.5 \Gamma_L$, $\Gamma_S = 6 \Gamma_L$ and $V_R = V_S = 0$. Notice that the DAR and CAR channels are dominating and they are responsible for the zero bias features.}\label{fig9}
\end{figure}

The contributions to the local conductance $\mathcal{G}_{L/L}^{TOT}$ are shown in Fig.~\ref{fig9} for strong coupling to the superconducting lead ($\Gamma_S = 6 \Gamma_L$) at temperature $k_B T = 0.01 \Gamma_L$, lower than the Kondo temperature $k_B T_K \approx 0.39 \Gamma_L$ evaluated from Eq.~(\ref{tkondo}). The zero bias enhancements of conductances are clearly visible in the DAR and CAR components. For the assumed values of parameters the direct Andreev reflection component $\mathcal{G}_{L/L}^{DAR}$ dominates close to $eV_L = 0$. It is a symmetric function of voltage applied to the left electrode. On the other hand the conductance due to the direct electron transfer between the normal electrodes is not so strongly influenced by the Kondo correlations. Increasing temperature suppresses the Abrikosov-Suhl resonance in the density of states and thereby has a detrimental effect on the zero bias anomaly in the conductance $\mathcal{G}_{L/L}^{TOT}$. The heights of other peaks change only slightly to accommodate the spectral weight of such vanishing peak.

One of our main findings is the appearance of negative non-local conductance $\mathcal{G}_{R/L}^{TOT}$ at zero bias as shown in lower panel of Fig.~\ref{fig9}. In the right electrode CAR and ET processes compete with each other and for all voltages, except close to $eV_L = 0$, the direct transfer dominates. Only around zero bias the CAR dominates. This is due to the increased effective transmission \textit{via} quantum dot due to the resonant state as it follows from the condition (\ref{factor}). Due to strong energy dependence of the self-energy, the CAR contributions to the non-local conductance dominate only in the vicinity of the Kondo resonance. In this case the collective many body state~\cite{hewson1997} is responsible for the effect. Similar behavior related to the increase of the effective transmittance has been previously observed in studies of different tunnel structures~\cite{soller2011a, levy-yeyati2005, melin2004} in high transparency limit.

To get the information about the interplay between Andreev and Kondo effects in the half-filled dot limit we use the IPT approach. This approximation to the self-energy is known to give correct results for the density of states~\cite{kajueter1996} and the linear transport coefficients. In the spirit of the previous approximation (\ref{U_correction}) we calculate diagonal self energy. In this approach the self-energy is chosen in such a way that it properly interpolates~\cite{martinrodero2011, Yamada-11, levy-yeyati1993} between exact second order in $U$ perturbative and the atomic limit formulas and has correct high frequency behavior.

In the 'superconducting atomic limit' the energy gap $\Delta$ exceeds the Kondo scale characterized by the Kondo temperature ($\Delta \gg k_B T_K$). This means no direct tunneling of electrons between the dot and superconducting electrode. Due to the proximity between the quantum dot and the superconducting electrode the empty and doubly occupied states on the dot are mixed and the transport proceeds via Andreev states as discussed in the Introduction.

\begin{figure}
\includegraphics[width=0.4\textwidth, clip]{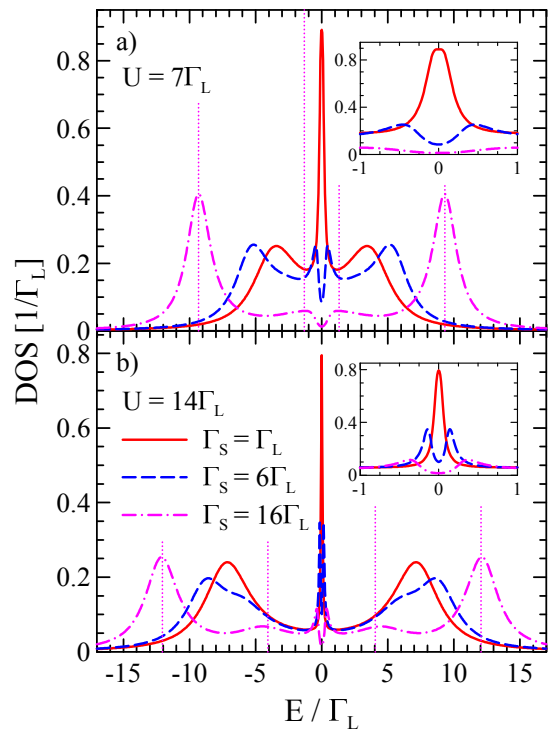}
\caption{(color online) The density of states of the correlated quantum dot for $\Gamma_S = \Gamma_L$ (red solid line), $\Gamma_S = 6 \Gamma_L$ (blue dashed line) and $\Gamma_S = 16 \Gamma_L$ (magenta dash-dot line) and (a) $U = 7 \Gamma_L$, $\epsilon_0 = -3.5 \Gamma_L$ (the estimated Kondo temperature $k_B T_K \approx 0.52 \Gamma_L$); (b) $U = 14 \Gamma_L$, $\epsilon_0 = -7 \Gamma_L$ (the estimated Kondo temperature $k_B T_K \approx 0.12 \Gamma_L$). The other parameters are $k_B T = 0.001 \Gamma_L$ and $\Gamma_R = 0.5 \Gamma_L$. In the insets the region around $E = 0$ are magnified. The dotted vertical lines indicate the positions of the Andreev bound states as calculated for Hubbard I approximation.}
\label{fig10}
\end{figure}

The tendency of the system to induce the superconducting correlations and the energy gap in the dot spectrum competes with the formation of the Abrikosov-Suhl resonance at the Fermi level. This resonance is a result of coupling to the normal leads and screening of the dot spin by spins of electrons in the conduction leads. The result of the competition obtained within IPT is shown in the Fig.~\ref{fig10} which presents the energy dependence of the dot density of states for half-filled case ($2 \epsilon_0 + U = 0$) for $U = 7 \Gamma_L$ and $U = 14 \Gamma_L$ and a few values of the couplings to the superconducting lead $\Gamma_S$.

In the Fig.~(\ref{fig10}) evolution of the subgap Andreev bound states is shown. Unfortunately in this approach the analytic expression for the bound state energies like (\ref{bound_states}) is not available. Nevertheless, in the figure we plot the positions of the bound states obtained from (\ref{bound_states}) as dotted lines and note good agreement with the positions of various features obtained from the numerical calculations, especially at high energies. This shows that the high energy spectrum undergoes small changes required to fulfill model independent sum rules, like that for the total number of states. At low energies and low temperatures $T$ the zero energy resonance  dominates the physics.

The central dip in the density of states (Fig.~\ref{fig10}) visible for large values of $\Gamma_S$ is related to the proximity induced pairing correlations on the dot. This feature disappears for a small $\Gamma_S \le \Gamma_L$. It is more pronounced for smaller values of $U$, when superconducting proximity effect dominates. For large values of the on-site repulsion (e.g. $U = 14 \Gamma_L$) four Andreev states are clearly visible for large coupling to the superconducting lead (e.g. $\Gamma_S = 16 \Gamma_L$). The insets to the Fig.~\ref{fig10} show the density of states close to the Fermi energy. The width of the unsplit Kondo resonance for $\Gamma_S = \Gamma_L$ depends on the correlation strength, being smaller for larger $U$. The increase of $\Gamma_S$ from $\Gamma_L$ to $6\Gamma_L$ results in the strong decrease of the Kondo peak accompanied by the apparent increase of the splitting.

Splitting of the Kondo resonance~\cite{yamada2010} is a characteristic feature of the half-filled quantum dot and usually disappears for $2 \epsilon_0 + U \ne 0$, similarly to the EOM results presented in Fig.~\ref{fig9}. It results from the interplay between the superconducting pairing and the Kondo singlet.

\section{Summary and Conclusions}

The contributions of various elementary transport processes to the currents flowing in a system consisting of the quantum dot contacted to one superconducting and two normal electrodes have been studied. The special attention was paid to the subgap local and the non-local Andreev type scattering events. For the non-interacting quantum dot and at $T = 0$ we obtained analytic expressions for differential conductances of all transport channels [Eqs.~(\ref{eq26})-(\ref{eq28})]. The main emphasis was on the influence of Coulomb interaction on the usual electron transfer (ET) between normal electrodes and the direct (DAR) and crossed (CAR) Andreev scattering and their interplay.

Treating the correlated quantum dot within the Hubbard I approximation (applicable for the description of the Coulomb blockade) we have numerically determined the effective energy spectrum and the differential conductances $\mathcal{G}_{L/L}$ and $\mathcal{G}_{R/L}$ for each transport channel. The ET processes have been shown to compete with the crossed Andreev reflections thereby limiting a possibility of obtaining the entangled electron pairs. The CAR processes dominate charge transport if the coupling to the superconducting electrode is much stronger than to the normal one. Coulomb interactions usually suppress the CAR conductances of the system except in the close vicinity of the Andreev bound states. The interplay between the direct and crossed Andreev reflections shows up both in local $\mathcal{G}_{L/L}$ and non-local $\mathcal{G}_{R/L}$ differential conductances.

To address the correlation effects in the Kondo regime we have used two complementary methods, based on the equation of motion procedure and the iterative perturbation theory. Itinerant electrons of the normal leads form the many body spin singlet state with electrons localized on the quantum dot. As a result, the narrow Abrikosov-Suhl resonance appears in the spectrum at the chemical potential for sufficiently low temperatures ($T < T_K$). This feature has a qualitative influence on the ET, DAR and CAR components of the conductance. All these transport channels reveal an enhancement of the low bias differential conductance, analogous to what have been observed experimentally in the metal-QD-superconductors junction~\cite{Deacon_etal}. The domination of the CAR processes in the non-local conductance of the Kondo correlated quantum dots is the most interesting finding. It shows that subtle quantum correlations (entanglement) between electrons forming Cooper pair are not destroyed by the formation of the many particle collective singlet states known as Kondo cloud.

In the Kondo regime the CAR is a dominant non-local transport channel at low voltages, leading to a negative value of the total zero bias conductance $\lim_{V_L \rightarrow 0}\mathcal{G}_{R/L}$. Let us note that the crucial role of interactions on the currents and current cross-correlations has also been found in the work on the hybrid devices with two quantum dots~\cite{rech2012}. Electron interactions which are expected to destroy quantum correlations in an electron gas in fact induce them in a suitably tuned nanodevices. In the three terminal system with all normal electrodes the Coulomb interactions lead to qualitative feedback effects showing up in the shot noise~\cite{bulka2008}.

It would be interesting to verify experimentally if the contributions $\mathcal{G}_{\alpha / \beta}^\kappa$ to the total differential conductance would indeed reveal the properties discussed in this paper. As the direct comparison of our results with the previous experiments~\cite{hofstetter2009, hermann2010, schindele2012} on the three terminal structures with two embedded quantum dots is impossible, we propose that the setup of Deacon et al.~\cite{Deacon_etal} with additional normal electrode could serve the purpose.

\section*{Acknowledgments:}

This work has been partially supported by the National Science Centre under the contract DEC-2011/01/B/ST3/04428 (KIW) and DEC-2012/05/B/ST3/03208 (GM, BRB), by the Ministry of Science and Higher Education grant No. N N202 2631 38 (TD) and by the EU project Marie Curie ITN NanoCTM (BRB, GM).


\begin{thebibliography}{11}

\bibitem{franceschi2010} S. D. Franceschi, L. Kouwenhoven, C. Sch\"{o}nenberger, and W. Wernsdorfer, Nature Nanotechnology \textbf{5}, 703 (2010).

\bibitem{eschrig2011} M. Eschrig, Physics Today \textbf{64}, 43 (January 2011).

\bibitem{burkard2000} G. Burkard, H. A. Engel, and D. Loss, Fortschr. Phys. \textbf{48}, 965 (2000); D. P. DiVincenzo, Fortschr. Phys. \textbf{48}, 771 (2000).

\bibitem{martinrodero2011} A. Mart\'{\i}n-Rodero and A. Levy Yeyati, Advances in Physics \textbf{60}, 899 (2011).

\bibitem{recher2001} P. Recher, E. V. Sukhorukov, and D. Loss, Phys. Rev. B \textbf{63}, 165314 (2001).

\bibitem{recher2002} P. Recher and D. Loss, Phys. Rev. B \textbf{65}, 165327 (2002).

\bibitem{beckmann2004} D. Beckmann, H. B. Weber, and H. v. L\"{o}hneysen, Phys. Rev. Lett. \textbf{93}, 197003 (2004).

\bibitem{russo2005} S. Russo, M. Kroug, T. M. Klapwijk and A. F. Morpurgo, Phys. Rev. Lett. \textbf{95}, 027002 (2005).

\bibitem{wei2010} J. Wei and V. Chandrasekhar, Nature Physics \textbf{6}, 494 (2010).

\bibitem{hofstetter2009} L. Hofstetter, S. Csonka, J. Nyg{\aa}rd, and C. Sch\"{o}nenberger, Nature \textbf{461}, 960 (2009).

\bibitem{hermann2010} L. G. Herrmann, F. Portier, P. Roche, A. L. Yeyati, T. Kontos, and C. Strunk, Phys. Rev. Lett. \textbf{104}, 026801 (2010).

\bibitem{lu2009} R. L\"{u}, H.-Z. Lu, X. Dai, and J. Hu, J. Phys.: Condens. Matter \textbf{21}, 495304 (2009).

\bibitem{futterer2009} D. Futterer, M. Governale, M. G. Pala, and J. K\"{o}nig, Phys. Rev. B \textbf{79}, 054505 (2009).

\bibitem{futterer2010} D. Futterer, M. Governale, and J. K\"{o}nig, EPL \textbf{91}, 47004 (2010).

\bibitem{eldridge2010} J. Eldridge, M. G. Pala, M. Governale, and J. K\"{o}nig, Phys. Rev. B \textbf{82}, 184507 (2010).

\bibitem{lambert1998} C. J. Lambert and R. Raimondi, J. Phys.: Condens. Matter \textbf{10}, 901 (1998); A. Kormanyos, I. Grace, and C. J. Lambert, Phys. Rev. B \textbf{79}, 075119 (2009).

\bibitem{martinrodero2012} A. Mart\'{\i}n-Rodero and A. Levy Yeyati, J. Phys.: Condens. Matter \textbf{24}, 385303 (2012).

\bibitem{whan1996} C. B. Whan and T. P. Orlando, Phys. Rev. B \textbf{54}, R5255 (1996).

\bibitem{levy1997} A. Levy Yeyati, J. C. Cuevas, A. L\'{o}pez-D\'{a}valos, and A. Mart\'{i}n-Rodero, Phys. Rev. B \textbf{55}, R6137 (1997).

\bibitem{kang1998} K. Kang, Phys. Rev. B \textbf{57}, 11891 (1998).

\bibitem{fazio1998} R. Fazio and R. Raimondi, Phys. Rev. Lett. \textbf{80}, 2913 (1998); \textit{ibid.} \textbf{82}, 4950 (1999).

\bibitem{schwab1999} P. Schwab and R. Raimondi, Phys. Rev. B \textbf{59}, 1637 (1999).

\bibitem{raimondi1999} R. Raimondi and P. Schwab, Superlatt. Mictrostruct. \textbf{25}, 141 (1999).

\bibitem{ivanov1999} T. I. Ivanov, Phys. Rev. B \textbf{59}, 169 (1999).

\bibitem{sun1999} Q.-F. Sun, J. Wang, and T.-H. Lin, Phys. Rev. B \textbf{59}, 3831 (1999); Phys. Rev. B \textbf{62}, 648 (2000); Y. Zhu, T.-H. Lin, and Q.-F. Sun, Phys. Rev. B \textbf{69}, 121302 (2004).

\bibitem{MKKIWActa} M. Krawiec and K. I. Wysoki\'{n}ski, Acta. Phys. Pol. A \textbf{97}, 197 (2000).

\bibitem{clerk2000} A. A. Clerk, V. Ambegaokar, and S. Hershfield, Phys. Rev. B \textbf{61}, 3555 (2000).

\bibitem{avishai2001} Y. Avishai, A. Golub, and A. D. Zaikin, Phys. Rev. B \textbf{63}, 134515 (2001).

\bibitem{sun2000} Q.-F. Sun, B.-G. Wang, J. Wang, and T.-H. Lin, Phys. Rev. B \textbf{61}, 4754 (2000).

\bibitem{krawiec2002} M. Krawiec and K. I. Wysoki\'{n}ski, Supercond. Sci. Technol. \textbf{17}, 103 (2002).

\bibitem{liu2004} S. Y. Liu and X. L. Lei, Phys. Rev. B \textbf{70}, 205339 (2004).

\bibitem{bergeret2006} F. S. Bergeret, A. Levy Yeyati, and A. Mart\'{\i}n-Rodero, Phys. Rev. B \textbf{74}, 132505 (2006); Phys. Rev. B \textbf{76}, 174510 (2007).

\bibitem{claughton1995} N. R. Claughton, M. Leadbeater, and C. J. Lambert, J. Phys.: Condens. Matter \textbf{7}, 8757 (1995).

\bibitem{tanaka2007} Y. Tanaka, N. Kawakami, and A. Oguri, J. Phys. Soc. Japan \textbf{76}, 074701 (2007) [see also \textbf{77}, 098001 (2008)]; Physica E \textbf{40}, 1618 (2008); Phys. Rev. B \textbf{78}, 035444 (2008); Phys. Rev. B \textbf{81}, 075404 (2010).

\bibitem{andersen2011} B. M. Andersen, K. Flensberg, V. Koerting, and J. Paaske, Phys. Rev. Lett. \textbf{107}, 256802 (2011); V. Koerting, B. M. Andersen, K. Flensberg, and J. Paaske, Phys. Rev. B \textbf{82}, 245108 (2010).

\bibitem{hiltscher2011} B. Hiltscher, M. Governale, J. Splettstoesser, and J. K\"{o}nig, Phys. Rev. B \textbf{84}, 155403 (2011).

\bibitem{buitelaar2002} M. R. Buitelaar, T. Nussbaumer, and C. Sch\"{o}nenberger, Phys. Rev. Lett. \textbf{89}, 256801 (2002); M. R. Buitelaar, W. Belzig, T. Nussbaumer, B. Babi\'c, C. Bruder, and C. Sch\"{o}nenberger, Phys. Rev. Lett. \textbf{91}, 057005 (2003).

\bibitem{cleuziou2006} J.-P. Cleuziou, W. Wernsdorfer, V. Bouchiat, T. Ondar\c{c}uhu, and M. Monthioux, Nature Nanotechnology \textbf{1}, 53 (2006).

\bibitem{jarrillo-herrero2006} P. Jarillo-Herrero, J. A. van Dam, and L. P. Kouwenhoven, Nature \textbf{439}, 953 (2006); J. A. van Dam, Y. V. Nazarov, E. P. A. M. Bakkers, S. De Franceschi, and L. P. Kouwenhoven, Nature \textbf{442}, 667 (2006).

\bibitem{jorgensen2006} H. I. Jorgensen, K. Grove-Rasmussen, T. Novotn\'{y}, K. Flensberg, and P. E. Lindelof, Phys. Rev. Lett. \textbf{96}, 207003 (2006); K. Grove-Rasmussen, H. I. J{\o}rgensen, and P. E. Lindelof, New J. Phys. \textbf{9}, 124 (2007); H. I. J{\o}rgensen, T. Novotn\'{y}, K. Grove-Rasmussen, K. Flensberg, and P. E. Lindelof, Nano Lett. \textbf{7}, 2441 (2007).

\bibitem{eichler2007} A. Eichler, M. Weiss, S. Oberholzer, C. Sch\"{o}nenberger, A. Levy Yeyati, J. C. Cuevas, and A. Mart\'{\i}n-Rodero, Phys. Rev. Lett. \textbf{99}, 126602 (2007); A. Eichler, R. Deblock, M. Weiss, C. Karrasch, V. Meden, C. Sch\"{o}nenberger, and H. Bouchiat, Phys. Rev. B \textbf{79}, 161407 (2009).

\bibitem{sand-jespersen2007} T. Sand-Jespersen, J. Paaske, B. M. Andersen, K. Grove-Rasmussen, H. I. J{\o}rgensen, M. Aagesen, C. B. S{\o}rensen, P. E. Lindelof, K. Flensberg, and J. Nyg{\aa}rd, Phys. Rev. Lett. \textbf{99}, 126603 (2007).

\bibitem{pillet2010} J. D. Pillet, C. H. L. Quay, P. Morfin, C. Bena, A. L. Yeyati, and P. Joez, Nature Physics \textbf{6}, 965 (2010).

\bibitem{meshke2011} M. Meschke, J. T. Peltonen, J. P. Pekola, and F. Giazotto, Phys. Rev. B \textbf{84}, 214514 (2011).

\bibitem{Deacon_etal} R. S. Deacon, Y. Tanaka, A. Oiwa, R. Sakano, K. Yoshida, K. Shibata, K. Hirakawa, and S. Tarucha, Phys. Rev. Lett. \textbf{104}, 076805 (2010); Phys. Rev. B \textbf{81}, 121308(R) (2010) and the supplemented on-line information.

\bibitem{schindele2012} J. Schindele, A. Baumgartner, and C. Sch\"{o}nenberger, Phys. Rev. Lett. \textbf{109}, 157002 (2012).

\bibitem{andreev1964} A. F. Andreev, Zh. Eksp. Teor. Fiz. \textbf{46}, 1823 (1964) [Engl. transl. Sov. Phys. JETP \textbf{19}, 1228 (1964)].

\bibitem{zubariev1960} D. N. Zubarev, Usp. Fiz. Nauk \textbf{71}, 71 (1960) [Engl. transl. Sov. Phys. Usp. \textbf{3}, 320 (1960)].

\bibitem{kajueter1996} H. Kajueter and G. Kotliar, Phys. Rev. Lett. \textbf{77}, 131 (1996).

\bibitem{levy-yeyati1993} A. Levy Yeyati, A. Mart\'{\i}n-Rodero, and F. Flores, Phys. Rev. Lett. \textbf{71}, 2991 (1993); J. C. Cuevas, A. Levy Yeyati, and A. Mart\'{\i}n-Rodero, Phys. Rev. B \textbf{63}, 094515 (2001).

\bibitem{meng2009} T. Meng, S. Florens, and P. Simon, Phys. Rev. B. \textbf{79}, 224521 (2009).

\bibitem{soller2011} H. Soller and A. Komnik, Physica E \textbf{44} 425 (2011).

\bibitem{pala2007} M. G. Pala, M. Governale, and J. K\"{o}nig, New J. Phys. \textbf{9}, 278 (2007) [see also New J. Phys. \textbf{10}, 099801 (2008)].

\bibitem{zhu2001} Y. Zhu, Q.-F. Sun and T.-H. Lin, Phys. Rev. B \textbf{65} 024516 (2001).

\bibitem{haug-jauho1996} H. Haug and A.-P. Jauho, \textit{Quantum Kinetics in Transport and Optics of Semiconductors, Second, Substantailly Revised Edition}, Springer Verlag, Berlin (2008).

\bibitem{Domanski-08} T. Doma\'{n}ski and A. Donabidowicz, Phys. Rev. B \textbf{78}, 073105 (2008); T. Doma\'{n}ski, A. Donabidowicz, and K. I. Wysoki\'{n}ski, Phys. Rev. B \textbf{78}, 144515 (2008); Phys. Rev. B \textbf{76}, 104514 (2007).

\bibitem{niu1999} C. Niu, D. L. Lin, and T.-H. Lin, J. Phys.: Condens. Matt. \textbf{11}, 1511 (1999).

\bibitem{meir1991} Y. Meir, N. S. Wingreen, and P. A. Lee, Phys. Rev. Lett. \textbf{66}, 3048 (1991); \textit{ibid.} \textbf{70}, 2601 (1993).

\bibitem{HubbardI} J. Hubbard, Proc. Roy. Soc. (London) A \textbf{281}, 401 (1964).

\bibitem{hewson1997} A. C. Hewson \textit{The Kondo problem problem to heavy fermions}, Cambridge studies in magnetism, Cambridge University Press (2007).

\bibitem{Glazman} L. I. Glazman and M. E. Raikh, Pis'ma Zh. Eksp. Teor. Fiz. \textbf{48}, 378 (1988) [Engl. transl. JETP Lett. \textbf{47}, 452 (1988)].

\bibitem{NgLee} T. K. Ng and P. A. Lee, Phys. Rev. Lett. \textbf{61}, 1768 (1988).

\bibitem{Yamada-11} Y. Yamada, Y. Tanaka, and N. Kawakami, Phys. Rev. B \textbf{84}, 075484 (2011).

\bibitem{kashcheyevs2006} V. Kashcheyevs, A. Aharony, and O. Entin-Wohlman, Phys. Rev. B \textbf{73}, 125338 (2006).

\bibitem{galperin2007} M. Galperin, A. Nitzan, and M. A. Ratner, Phys. Rev. B \textbf{76}, 035301 (2007).

\bibitem{krawiec2002a} M. Krawiec and K. I. Wysoki\'{n}ski, Phys. Rev. B \textbf{66}, 165408 (2002).

\bibitem{sellier2005} G. Sellier, T. Kopp, J. Kroha, and Y. S. Barash, Phys. Rev. B \textbf{72}, 174502 (2005).

\bibitem{glazman1989} L. I. Glazman and K. A. Matveev, JETP Lett. \textbf{49}, 659 (1989).

\bibitem{munoz2013} E. Munoz, C. J. Bolech, and S. Kirchner, Phys. Rev. Lett. \textbf{110}, 016601 (2013); S. Smirnov and M. Grifoni, Phys. Rev. B \textbf{87}, 121302(R) (2013).

\bibitem{Bauer_2013} A. Oguri, Y. Tanaka, and J. Bauer, Phys. Rev. B \textbf{87}, 075432 (2013).

\bibitem{soller2011a} H. Soller and A. Komnik, Eur. Phys. J. D \textbf{63}, 3 (2011).

\bibitem{levy-yeyati2005} A. Levy Yeyati, J. C. Cuevas, and A. Martin-Rodero, Phys. Rev. Lett. \textbf{95}, 056804 (2005).

\bibitem{melin2004} R. Melin and D. Feinberg, Phys. Rev. B \textbf{70}, 174509 (2004).

\bibitem{yamada2010} Y. Yamada, Y. Tanaka, and N. Kawakami, Physica C \textbf{470}, S875 (2010).

\bibitem{rech2012} J. Rech, D. Chevallier, T. Jonckheere, and T. Martin, Phys. Rev. B \textbf{85}, 035419 (2012).

\bibitem{bulka2008} B. R. Bu{\l}ka, Phys. Rev. B \textbf{77}, 165401 (2008).

\end{thebibliography}
\end{document}